# Precision mass measurements of neutron-rich nuclei between N=50 and 82

Juha Äystö

Department of Physics, University of Jyväskylä, Finland
Helsinki Institute of Physics, University of Helsinki, Finland
E-mail: juha.aysto@helsinki.fi

**Abstract.** Our knowledge of binding energies of neutron-rich nuclei has experienced a major revision during the last five years due to the introduction of Penning-trap based mass measurements. New mass values for nearly 300 nuclides produced in fission with uncertainties of 10 keV or less have become available. The data produced at three Penning trap facilities at Jyvaskyla, CERN-ISOLDE and Argonne cover all isotopic chains from Ni to Pr, except iodine. In this talk some of this data is reviewed and applied using the mass differentials such as two-neutron binding energy and odd-even staggering to probe their sensitivity on changes in nuclear structure and on the strength of the N=82 shell gap and associated pairing effects.

## 1. Introduction

Nuclear masses are the anchor points which define the absolute reference frame for nuclear structure physics. The total mass of a nucleus is defined by the sum of the nucleon masses and their total binding energy. The evolution of the nuclear mass or binding energy surface as a function of proton and neutron numbers is determined by the nucleonic many-body quantum mechanical system. Interplay of collective behaviour and shell-like structure of valence nucleon energy states coupled with the pairing interaction set also the limits for the existence of bound nuclei in determining the nucleon drip lines as well as the stability of the heaviest elements. At best, the total mass energy of a nucleus can presently be calculated with accuracy of the order of a few hundred keV which corresponds to a relative mass uncertainty between $10^{-5}$ and $10^{-6}$. Such accuracies have been routinely available by standard time-of-flight, reaction Q-value and decay energy measurements. In addition to the absolute mass or binding energy, the first and second order differentials of masses are important indicators of local and global changes in structures offering a challenge for precision mass measurements of nuclear ground states and low-energy isomeric states.

In general, the mass measurements of neutron-rich nuclei far from stability require a relative precision of $\delta m/m$ less than $10^{-6}$, corresponding to <100 keV at A = 100, for probing the evolution of shell structure, collectivity and shell closures [1,2]. These can be probed by the systematics of one- and two-nucleon separation energies over long isotopic or isotonic chains. A higher accuracy of $\delta m/m < 10^{-7}$ (e.g. 10 keV at A = 100) is required in measuring the binding energies of highly neutron-rich nuclei with a halo structure. Relative mass accuracies of the order of $10^{-8}$ are needed in studies of second-order mass differentials, such as odd–even staggering which is connected to the nature of pairing in neutron-rich nuclei [3], including its isospin dependence and impact in position of the neutron drip-line. Similar relative mass accuracy of $10^{-8}$ is required for a studying specific class of double mass differences δV pn [4], which has been found to possess some correlation with local effects such as octupole correlations in nuclear structure [5].

Mass measurement techniques based on frequency measurements of ions in storage rings or ion traps have been developed to their full capacity in just during the last ten years or so, see ref. [6] and refs. therein. In particular, introduction of Penning traps to radioactive isotope mass measurements has changed the scope of the direct mass measurements significantly. Fast buffer-gas stopping and cooling of low-energy ions using a radiofrequency driven multipole trap has made possible very fast injection of ions into the Penning trap. The first generation traps were connected to traditional ISOL systems, such as ISOLDE at CERN, where slow extraction, especially, for refractory elements limited the applicability of the method to a few chemical elements. This problem has now been solved in the IGISOL technique and its successor technique called the ion-catcher method. The combination of the above mentioned approaches have opened up a possibility for high-precision mass measurements of radioactive isotopes of all chemical elements with half-lives down to less than 100 ms. Figure 1 shows all neutron-deficient and neutron-rich isotopes whose masses have been measured with JYFLTRAP, see *http://research.jyu.fi/igisol/JYFLTRAP_masses*. These isotopes were produced in fusion and fission reactions using the IGSOL facility coupled to the K=130 MeV cyclotron of the JYFL accelerator laboratory.

In this talk, an overview of recent Penning-trap mass measurements of neutron-rich nuclei produced in fission reactions is given and their implication for nuclear structure physics is discussed when relative mass accuracies of the order of $10^{-8}$ are available. A more detailed review is given in ref. [7].

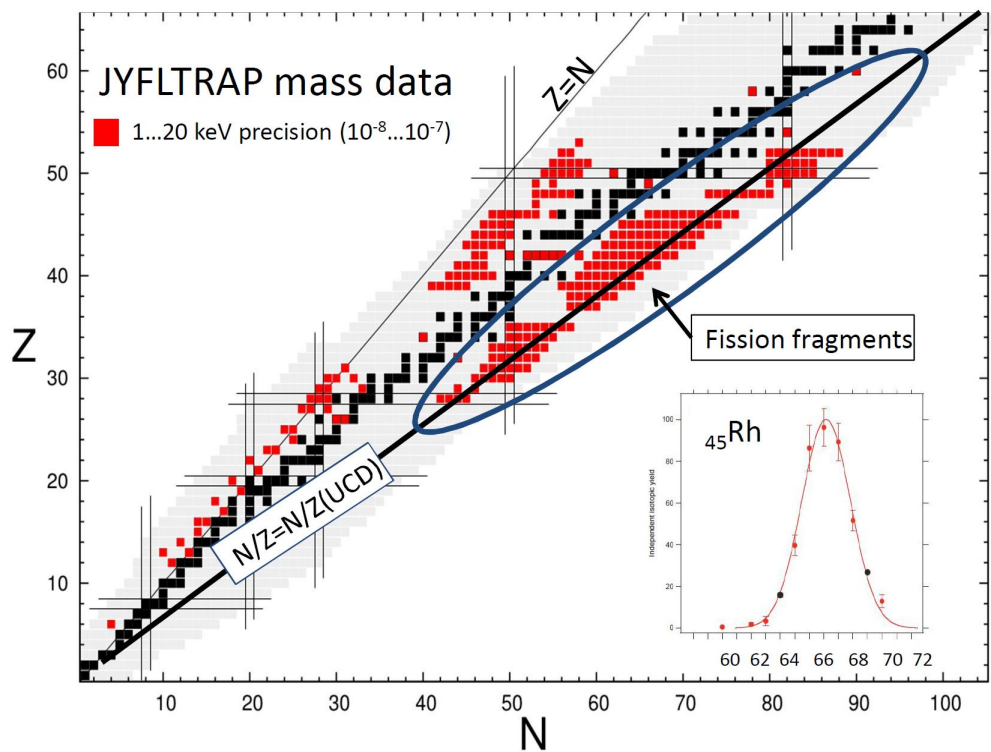

Figure 1. Summary of isotopes whose masses have been measured by JYFLTRAP coupled to the IGISOL mass separator facility. UCD is unchanged charge distribution following fission of $^{238}$U. The inset labeled "Rh" denotes the measured independent yield distribution of Rh isotopes.

## 2. Experimental methods

Several recent papers have presented various experimental approaches where the Penning traps have been employed for accurate mass determinations of neutron-rich isotopes. In this paper we focus our discussion on ISOL-based mass spectrometry of fission products. Two ISOL facilities employing this principle are the traditional thick target facilities ISOLDE at CERN and TITAN at ISAC in TRIUMF. Another type of ISOL facilities are those employing a gas catcher for stopping fission fragments before their extraction for mass selection and further analysis. The setup used in Jyväskylä is shown in Figure 2. Singly charged fission product ions from proton-induced reactions on a natural uranium target are produced and selected according to their mass by IGISOL and are then injected into the JYFLTRAP system. Due to very short delay time of the order of milliseconds of the method, separated ions are primary fission fragments in their ground or isomeric states. Following the mass(number) separation the ions forming a DC-like beam are injected into a buffer-gas filled RFQ trap for cooling and bunching. Ions are then injected as bunches of a few μs long to the double Penning trap system operating in a 7 Tesla magnetic field. The first trap is called a purification trap that provides mass-selective buffer gas cooling with a typical mass resolving power $m/\Delta m \leq 10^5$. This resolution can easily provide clean separation of neighbouring isobars before injecting them into the second trap, termed a precision trap.

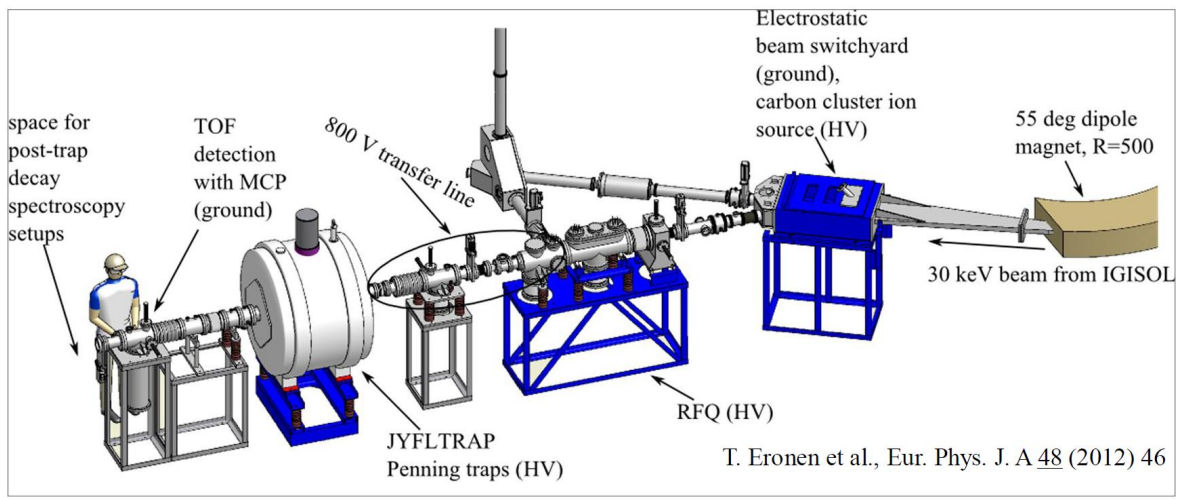

Figure 2. JYFLTRAP setup consisting of the RFQ buncher and coller and a double Penning trap.

In a Penning trap ions have three different eigenmotions, one axial motion and two radial motions consisting of a slow magnetron motion ($\nu_+$) and fast reduced cyclotron motion ($\nu_-$). These two frequencies sum up to a true cyclotron frequency with high precision even in a perfectly non-ideal Penning trap. True cyclotron frequency is dependent on the mass of an ion as follows:

$$\nu_c = qB/(2\pi m)$$

where B is he magnetic field and q and m are the charge and the mass of the ion. Therefore, the mass of an ion is determined by by measuring its cyclotron frequency against the cyclotron frequency of another ion with a well-known mass. Typically, relative mass accuracy of the order of $10^{-7}$ to $10^{-8}$ can be reached. The cyclotron frequency is determined using resonant quadrupolar RF excitation to excite the ions' radial energy. This energy can be measured either by a pick-up signal or by time-of-flight technique as shown in the bottom part of Figure 2. The detailed description of this procedure is given in reference [8].

In an ideal case only one ion at a time is needed for a measurement of the cyclotron frequency. Therefore, the method is applicable for relatively low production rates. At IGISOL typical highest independent nuclide yields have been measured to be $10^5$ ions/s. With the minimum required production rate of a few ions/s about 400 short-lived neutron-rich isotopes are within reach for the mass measurement experiments with JYFLTRAP.

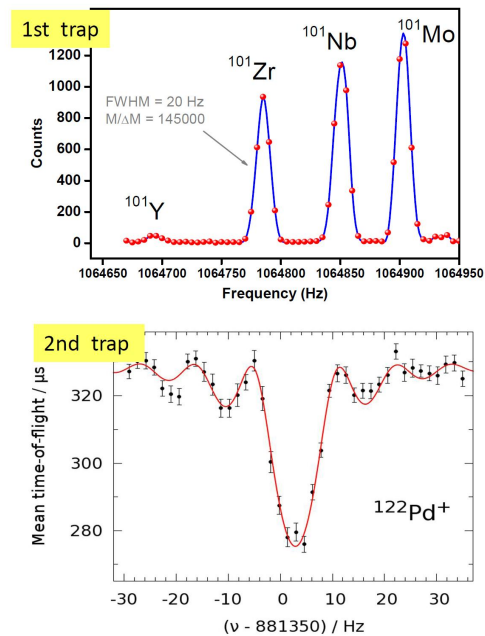

Figure 3. (Top) Mass spectrum of a selected range of A=101 fission product isobars. (Bottom) Time-of-flight spectrum over the true cyclotron resonance region for $^{122}$Pd ions.

## 3. Summary of results

During recent years, masses of over 300 fission fragments have been measured with Penning traps. A detailed list of references for these studies is given in ref. [7]. Until now, most of the mass data have been obtained using the JYFLTRAP setup. First direct mass measurements of radioactive (neutron-rich) isotopes were done at ISOLDE using the ISOLTRAP facility. Some nuclei in the fission product region have also been studied with the Canadian Penning Trap CPT at ANL and LEBIT at MSU. The data on masses measured at JYFLTRAP and ISOLTRAP are available on the websites at [*http://research.jyu.fi/igisol/JYFLTRAP_masses,http://isoltrap.web.cern.ch/isoltrap/database/isodb.asp*].

For fission products a majority of earlier mass data before the Penning-traps became available were derived from beta-decay endpoint energies. As an indirect method this approach tends often to underestimate the mass value because of a limited knowledge of the daughter states and unobserved feedings to high-lying excited states. Mostly for this reason the direct mass measurements by Penning traps have revealed large systematic discrepancies compared to the tabulated masses. The effect becomes pronounced towards the neutron-rich side of an isobaric chain. It is clearly seen in Figure 3 where the difference between the Penning trap based values of fission fragments and the Atomic Mass Evaluation 2003 [9] values has been plotted as a function of the distance from the stability line. This line is approximated by the formula $N_{stab}=(Z*1,64-12)$. Some of the earliest ISOLTRAP mass values have already been included in the AME03. Therefore, they deviate less from the AME03 than the other experimental results. For readability the error bars due to the AME03values are not shown in figure. Typically, they are less than 100 keV except for the extrapolated values where the unceratinties given are of the order of 500 keV. It is obvious that the systematic uncertainties of the "old data" were often severely underestimated. A general conclusion drawn is that neutron-rich nuclei seem to be less bound than judging from the "old" data.Therefore, the comparison to the values from the mass model calculations may have led to misleading conclusions.

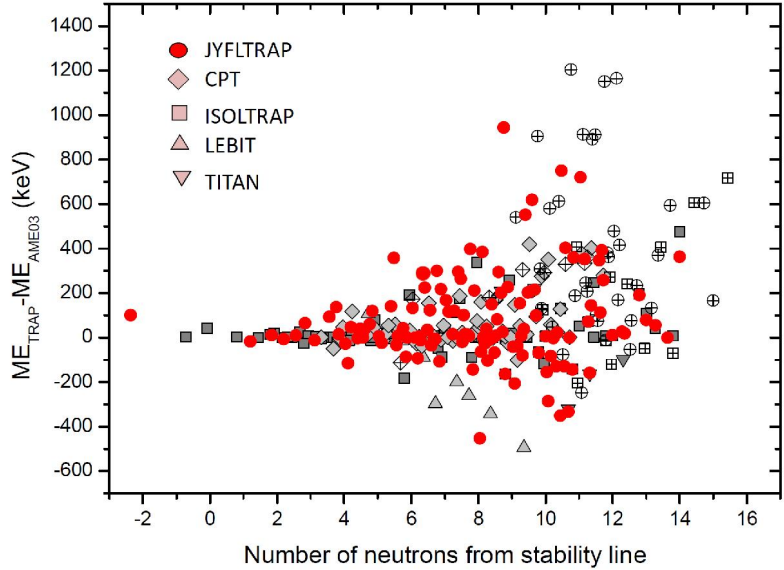

Figure 4. Difference between the new Penning trap data and the AME03 tabulated masses. The crossed symbols are for the extrapolated values from AME03.

The forthcoming new Atomic Mass Evaluation 2012 [10] presents a substantially improved data set. Most of the new results have been included in it except for the latest results from JYFLTRAP [11, 12]. Nevertheless, these new results deviate less from the AME11 values than the Penning-trap measurements in general from the AME03. Also, the results of the refs. [11, 12] are in agreement with the extrapolated values of AME11 within their uncertainties. There are only a few cases where the Penning trap data show more than $1\sigma$ deviation from the AME11evaluation, see ref. [7].

## 4. Discussion

Understanding the evolution the shell structure and collective properties as a function of proton and neutron numbers requires accurate knowledge of the fine structure of the mass surface. They can be studied by the systematic of mass differentials as a function of proton and neutron numbers. Most often used differentials are one- and two-nucleon separation energies and decay Q-values. Examples of useful second order mass differentials are the shell gap energies and odd-even staggering of masses which are sensitive to magic nucleon numbers and pairing effects, respectively. With the present-day ion-trap spectrometry these quantities are typically available with accuracies of the order of 10 keV or better. This offers accuracy comparable to that of excited states spectroscopy far from stability in the outskirts of the known nuclear landscape, and provides a true challenge and opportunity for mass models and structure theories.

### 4.1. Two-neutron binding energies

Two-neutron separation energy $S_{2n}$ of a nucleus with Z protons and N neutrons can be obtained by using the following formula:

$$S_{2n} = E(A,Z) - E(A-2,Z) = [M(A-2,Z) + 2*M_n - M(A,Z)]c^2,$$

where E and $Mc^2$ stands for the binding energy and mass, respectively. In the following selected two-neutron separation energy systematics are presented and discussed. They focus in two regions of neutron-rich nuclei; the deformed region with N=60 and the doubly magic $^{132}$Sn region. Instead of viewing the behavior of two-neutron separation energies in a traditional way as a function of the neutron number we choose to plot them as a function of the proton number for adjacent isotope chains. This provides another perspective on the evolution of shell gaps far from stability.

*Deformed region around A=100*

Region of neutron-rich nuclei between N=50 and 82 presents a landscape of different structures which also impact the mass-energy. In particular, the neutron-rich isotopes around A ≈ 100 have been found to possess shape transitions and coexistence of shapes around Z = 40 and N = 60, see [13] and references therein. While the ground states of the strontium and zirconium isotopes below N ≈ 60 appear to be only weakly deformed or nearly spherical, the heavier isotopes display mainly axially symmetric deformed shapes. Spectroscopic studies have shown that nuclei with N > 60 have large ground state quadrupole deformations, while the intermediate N = 59 isotones of strontium, yttrium and zirconium still have nearly spherical ground states, see ref. [14]. This interpretation has since then been confirmed by a series of collinear laser spectroscopy experiments in the form of a sudden increase of the mean-square charge radii around N = 60 [15]. Higher Z molybdenum isotopes are, however, found to be already less deformed or even possess triaxial shapes.

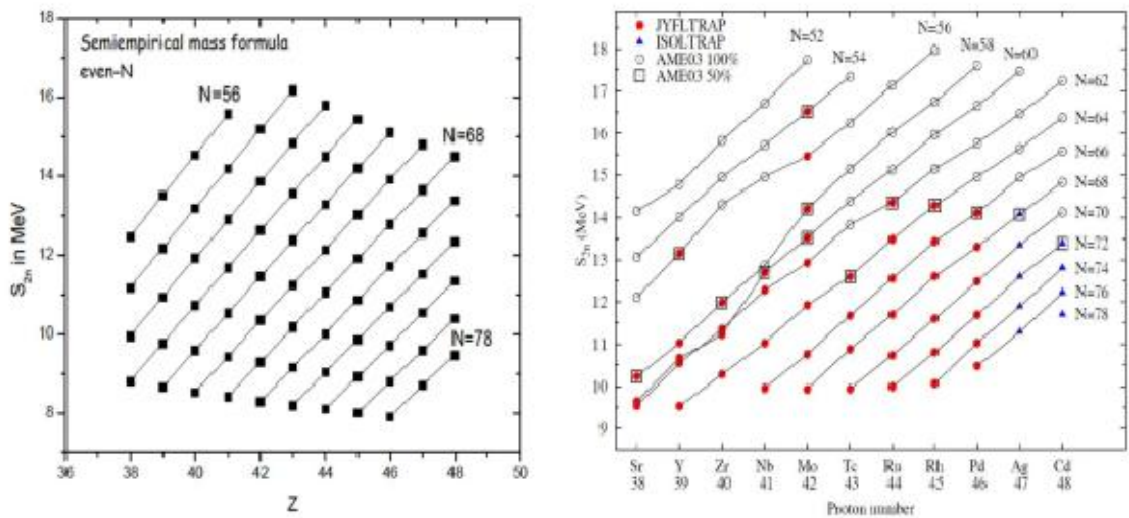

Figure 5. Two-neutron binding energy vs. Z as calculated with the liquid drop model (left) and as obtained from the recent experimental data (right). The uncertainties of the data points are within the size of each symbol.

In order to probe the impact of the changing deformation on ground states we have plotted $S_{2n}$ values for even isotones as function of the proton number in Fig. 5 for neutron-rich nuclei studied with the JYFLTRAP and ISOLTRAP. A distinct energy gap is observed when crossing the N = 56 neutron number at the semi doubly magic $^{96}$Zr preceding the rapid onset of deformation above N=58. For Zr one observes even a crossing of the curves corresponding to N=58 and 60 which represents an exceptional case not observed anywhere else in the nuclear chart. Otherwise, the curves show a smooth nearly monotonic behavior indicating only weak structural effects over a broad range of proton and neutron numbers. For a reference, it is interesting to view the measured values against a simple liquid drop model based on the "structureless" semiempirical mass formula. Using the coefficients given in ref. [16] two-neutron separation energies are plotted with the experimental values in the left part of Fig. 5. Except for the anomaly around N = 60 the overall trend in the slopes for Zr and Mo isotope chains are reproduced. However, the distance between the adjacent even-N isotones is nearly constant in the semiempirical model whereas it decreases for the experimental values implying an influence of the changes in shell structure and level density.

*The doubly-magic $^{132}$Sn region*

The shell closure at Z=50 and N=82 has long been known to exhibit features of exceptional purity for its spherical single particle structure [17]. However, only recently the accurate direct measurements of masses with Penning traps have become possible improving and extending our knowledge of masses and binding energies beyond N=82 from Sn to Xe. Most recent results obtained by the JYFLTRAP and ISOLTRAP facilities covered the masses of isotopic chains up to $^{135}$Sn, $^{136}$Sb, $^{140}$Te and $^{146}$Xe [5,11]. The summary of the newest measured masses is shown in Figure 6. Our present mass data provide no information on the behavior of the N = 82 shell closure below Sn which remains major challenge for future experiments. On the other hand, the Penning trap measurements at CPT have covered also heavier fission products and their masses from Ba up to Gd [19].

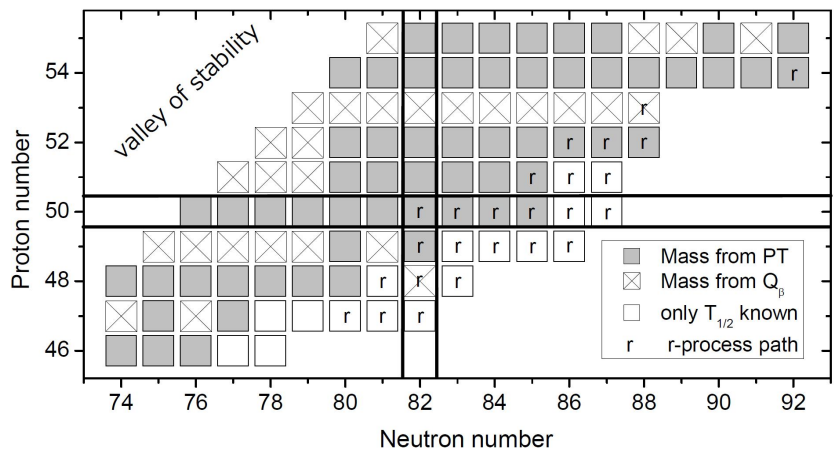

Figure 6. Neutron-rich isotopes around Z=50 and N=82 whose masses were recently measured with Penning traps. The r-process path is from ref. [18]. See text for other details and references.

One of the goals of today's nuclear structure research is to study the survival of magic shell closures far out of the valley of stability. Two-neutron separation energy across the N=82 shell closure shows a large and distinct drop of about 6.5 MeV from N=82 to N=84 at Z=50. This behavior is typical and is observed for crossings of all major closed shells near the valley of stability. Above Z=50 the value of this energy gap is decreasing with the increasing proton number indicating the importance of correlations induced by interplay between single particle and collective effects. The dependence of the shell gap energy $\Delta = S_{2n}(N = 82) - S_{2n}(N = 84)$ on the proton number given in figure 7 indicates the persistence of the N = 82 shell closure but shows a reduction of its value from 6.5 MeV to about 4 MeV in the middle of the proton shell.

This figure shows also theoretical values calculated with the mean-field models of Bender et al [2] and Goriely et al [20]. The calculation of Bender et al employs SLy4 [21] energy density functional (DFT) in the deformed basis with dynamical quadrupole correlations. A relatively good agreement with the experimental values is obtained although the model is not able to explain the reduction sufficiently, unlike in the case of N = 50 [22]. The same calculation without correlations gives a rather constant value as a function of proton number confirming the importance of correlations due to core polarization effects. Two other mean-field approaches of Goriely et al [20,23] employ the HFB framework either with a Gogny interaction (HFB-D1M) or a Skyrme force (HFB-21) taking into account all quadrupole correlations self-consistently and microscopically. This model seems provide reasonably good overall agreement. In conclusion, it is obvious that the mean-field models based on Skyrme interaction such as those in [2, 23] rather successfully describe the binding energies near the shell closure where over-prediction of the gap is only about 0.5 MeV.

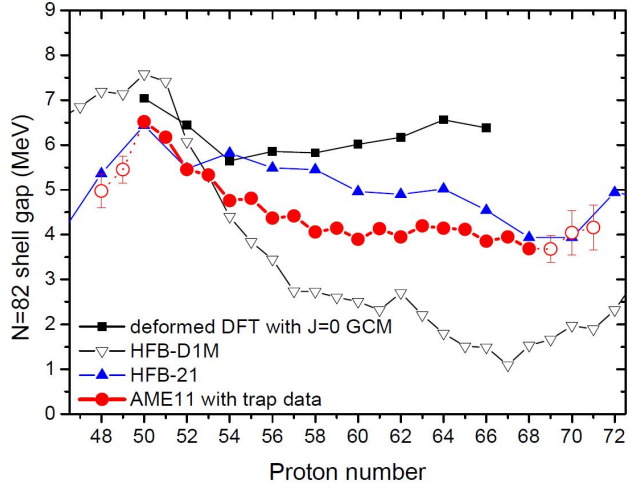

Figure 7. Two neutron shell gap $\Delta = S_{2n}(N = 82) - S_{2n}(N = 84)$ as a function of proton number. For details, see text.

*4.2. Odd-even staggering of masses and pairing gap*
High-accuracy in mass measurements opens new possibilities for systematic studies of structure effects in nuclear binding energies. An excellent example is the study of pairing effects. Pairing energy represents typically a relative contribution to a total mass energy of the order of $10^{-5}$ to $10^{-6}$. The role of pairing energy is particularly importance in weakly bound nuclei far from the valley of stability and near the nucleon drip lines. Therefore, their formulation is an important task for nuclear theory. Empirically, pairing energies can be studied by measuring odd-even mass differences along the isotopic or isotonic chains. Odd-even staggering (OES) has been largely attributed to BCS pairing but there are also other mechanisms, such as those due to mean-field effects, that can contribute (see e.g. [3, 24].

The simplest approach to study pairing in nuclei is a three-point odd-even mass staggering formula, which can be written for neutrons as:

$$\Delta^3(N) = (-1)^N [ME(Z,N+1) - 2ME(Z,N) + ME(Z,N-1)]/2$$

where ME is mass excess and Z and N denote proton and neutron numbers. One advantage of this formula is that it can be applied to more experimental data than the higher order formulae. The 3-point formula presents contributions from both pairing and mean-field effects. The odd-neutron values of $\Delta^3(N)$ can be considered to be a measure of pairing effects only whereas the even-N values are more sensitive to the splitting of the single-particle spectrum near the Fermi level [3]. Figure 8 (top) shows the measured odd-even staggering for Sn, Te and Xe isotopes between N= 71 and 91. To emphasize pairing effect, for each isotopic chain the points with odd neutron numbers are connected by solid lines. There is an obvious asymmetry across N = 82 but the behavior is rather different for Sn, Te, and Xe isotopes. Asymmetry is very distinct for Sn but considerably less for Te and Xe. This may indicate increasing importance of collective effects with increasing proton number above Sn.

In order to probe this question theoretically self-consistent calculations have been recently performed by Dobaczewski and Kortelainen [see ref. 4]
using the Sly4 [21] energy density functional and contact pairing force. The pairing channel was described within the HFB approximation and the blocking and filling approximations were used to treat odd nuclei. Figure 8 (bottom) shows the neutron odd-even staggering calculated within the spherical approximation [25]. The experimental decrease of OES when crossing the N = 82 gap for Sn is very well reproduced with the volume-dominated mixed pairing forces [26], whereas the data exclude the pure surface-localized pairing force. However, in the N = 83 isotones, the disagreement with data for Te and Xe remains a puzzle, unresolved within the current state of the art theoretical approaches. It cannot be explained by the combined pairing and static-deformation correlation effects. Study of higher-order correlations, such as the configuration mixing of deformed states, requires an implementation of methods for odd nuclei that are presently not available.

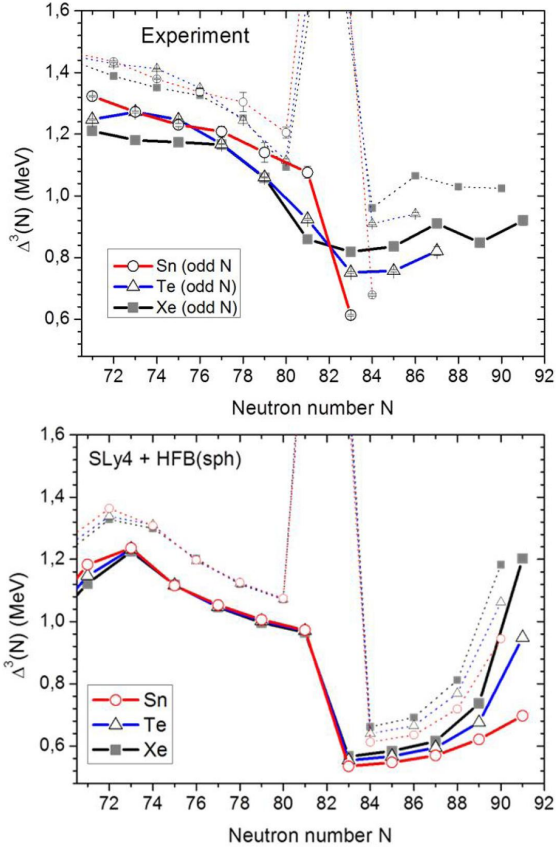

Figure. 8. Experimental (top) and theoretical (bottom) odd-even staggering for Sn, Te and Xe isotopes. For clarity, the odd-N points are connected by the solid line. The data points for N=82 are beyond the scale. For the definition of $\Delta^3$, see text.

*Conclusions*

Current status of accurate mass measurements with Penning traps has been reviewed and possible significance of earlier unavailable accuracies of the new mass data are viewed from the point of view of nuclear structure physics.

*Acknowledgements*

The author is indebted to the JYFLTRAP team and a large number of other colleagues who have contributed to obtaining the results reviewed in this talk. In particular, I would like to thank Ari Jokinen, Tommi Eronen, Anu Kankainen, Veli Kolhinen and Jacek Dobaczewski for numerous discussions and their impact on this work.

This work has been supported by the Academy of Finland under the Finnish Centre of Excellence Programme 2006-2011 (Nuclear and Accelerator Based Physics Research at JYFL).